\documentclass[a4,12pt]{article}
\usepackage{graphicx}
\usepackage[comma,authoryear]{natbib}
\usepackage[below]{placeins}
\usepackage{algorithm}
\usepackage[noend]{algpseudocode}
\usepackage{amsmath,amssymb,booktabs,bm,mm}
\usepackage{amsthm}
\usepackage{rotating}
\usepackage{hyperref}

\def\HD{\mathop{\mathrm{HD}}}
\def\nHD{\mathop{\mathrm{nHD}}}
\def\sp{\mathop{\mathrm{span}}}
\def\argmin{\mathop{\mathrm{arg\,min}}}
\def\forwhich{\,|\,}
\def\union{\cup}
\def\intersection{\cap}

\newtheorem{lemma}{Lemma}
\newtheorem{proposition}{Proposition}
\newtheorem{theorem}{Theorem}
\newtheorem{corollary}{Corollary}

\begin{document}

\title{Exact computation of the halfspace depth}

\author{Rainer Dyckerhoff \qquad Pavlo Mozharovskyi \\
{\small \indent} \\
{\small Institute of Econometrics and Statistics, University of Cologne} \\
{\small Albertus-Magnus-Platz, 50923 Cologne, Germany} \\
\indent\\
}

\date{January 12, 2016}

\maketitle

\begin{abstract}
For computing the exact value of the halfspace depth of a point
w.r.t. a data cloud
of $n$ points in arbitrary
dimension, a theoretical framework is suggested. Based on this framework a whole class of
algorithms can be derived. In all of these algorithms the depth is calculated
as the minimum over a finite number of depth values w.r.t. proper projections
of the data cloud. Three variants of this class are studied in more
detail. All of these algorithms are capable of dealing with data that are not
in general position and even with data that contain ties.
As is shown by simulations, all proposed algorithms prove to be very efficient.
\end{abstract}


\section{Introduction}

In 1975 John W. Tukey suggested a novel way of ordering multivariate data.
He proposed to order the data according to their centrality in the data
cloud. To achieve this, he defined what is nowadays known as \emph{halfspace
depth} (also called Tukey depth or location depth). Motivated by his proposal
many notions of data depth have been proposed in the last decades, e.g., the
simplicial depth \citep{Liu88,Liu90}, the projection depth
\citetext{\citealp{Liu92,ZuoS00,Zuo03}; based on a notion of 
outlyingness proposed by \citealp{Stahel81,Donoho82}}, and the zonoid depth 
\citep{KoshevoyM97}.                 

Data depths have been applied in many fields of statistics,
among others in multivariate data analysis \citep{LiuPS99}, statistical quality
control \citep{LiuS93},
classification \citep{MoslerH06,LangeMM14},                                                
tests for multivariate location and scale \citep{Liu92,Dyckerhoff02},
multivariate risk measurement \citep{CascosM07}, and robust linear programming
\citep{MoslerB14}.

Consider a data cloud $\bmX=(\bmx_1,\dots,\bmx_n)$ with data points $\bmx_i\in
\mathbb{R}^d$.
Conditioned on it, a statistical depth function assigns to an arbitrary point
$\bmz\in\mathbb{R}^d$ its degree of centrality, $\bmz\mapsto D(\bmz|\bmX)\in[0,1]$.

The halfspace depth is determined as the smallest fraction of data points contained
in a closed halfspace containing $\bmz$. This classical depth function,
based on the ideas of \citet{Tukey75} and later developed by~\citet{DonohoG92},                
is one of the most important depth notions and is historically the first one.
It possesses a number of attractive properties, such as affine invariance,
monotonicity on rays from any deepest point, quasiconcavity and upper
semicontinuity, studied in \citet{ZuoS00,Dyckerhoff04,Mosler13}.
The halfspace depth determines uniquely the empirical distribution~\citep{Koshevoy02}
and takes a finite number of values in the interval from 0 (for the points that
lie beyond the convex hull of the data) to its maximum value (which
depends on the data, but is $1/2$ if all data points are different from $\bmz$),                                        
increasing by a multiple of $1/n$.
By its nature the halfspace induced 
maximizers (the Tukey median) have a relatively high breakdown point.

The task of calculating the halfspace depth has a direct connection to the
regression depth of~\citet{RousseeuwH99} and to the risk-minimizing separating
hyperplane in classification.
Further, computing the halfspace depth essentially coincides with the
densest hemisphere problem \citep{JohnsonP78}, which is of non-polynomial
complexity in $(n,d)$.
For this reason a great part of the literature on the halfspace depth considers its computational aspects.
Below, we give an overview of the history of its exact (in Section~\ref{ssec:chronicles}) and approximate (in Section~\ref{ssec:approx}) computation.

\subsection{Previous approaches to calculation of the halfspace
depth}\label{ssec:chronicles} The idea of the halfspace depth has been introduced
in a conference paper by~\citet{Tukey75}. A similar mechanism of cutting by
hyperplanes (lines in the two-dimensional case) has also been used for a bivariate
sign test by~\citet{Hodges55}. During the last decades, a variety of
attempts have been made to compute the halfspace depth and its trimmed regions.

\citet{RousseeuwR96} pioneered in exactly calculating the half\-space depth for
bivariate data clouds and constructing its contours \citep{RutsR96a,RutsR96b} 
by exploiting the idea of a circular
sequence~\citep{Edelsbrunner87}. Here, the depth of a point is computed with
complexity $O(n\log{n})$, and a single depth region is constructed with
complexity $O(n^2\log{n})$, both essentially determined by the complexity of the
QUICKSORT procedure. \citet{JohnsonKN98} suggest to account for a small subset
of points only when constructing the first $l$ depth contours, which yields a
better complexity for small $l$ (algorithm FDC). The halfspace depth describes a
data cloud by a finite number of depth contours. \citet{MillerRRSSSS03} compute
all these with complexity $O(n^2)$, by which the depth of a single point can be
afterwards calculated with complexity $O(\log^2{n})$.

\citet{RousseeuwS98} introduce an algorithm to compute the halfspace depth for $d=3$
with complexity $O(n^2\log{n})$. They project points onto planes orthogonal
to the lines connecting each of the points from $\bmX$ with $\bmz$ and then
calculate the halfspace depth in these planes using the algorithm
of~\citet{RousseeuwR96}. \citet{BremnerFR06} calculate the halfspace depth with
a primal-dual algorithm by successively updating upper and lower bounds by means
of a heuristic till they coincide. \citet{BremnerCILM08} design an output-sensitive
depth-calculating algorithm that represents the task as two maximum subsystem problems
for $d>2$. The latter ones are then run in parallel.

An interesting issue of updating the depth when points are added continuously to
the data set is handled by \citet{BurrRS11} for bivariate depth and depth
contours.

\citet{KongM12} employ direction quantiles defined as halfspaces corresponding
to quantiles on univariate projections and prove their envelope to coincide with
the corresponding halfspace depth trimmed region.
In exact computation of depth trimmed regions for $d>2$, \citet{MoslerLB09}
pioneered for the zonoid depth \citep[see also][]{Mosler02}.
Their idea
to segment $\mathbb{R}^d$ into direction cones
has been exploited in later algorithms for computing depth and depth regions,
also
for the halfspace depth.
\citet{HallinPS10} establish a
direct connection between multivariate quantile regions and halfspace depth trimmed
regions. The multivariate directional quantile for a given direction corresponds to
a hyperplane that may carry a facet of a depth trimmed region. More than that,
the authors define a polyhedral cone containing all directions yielding the same
hyperplane; the union of the finite set of all these cones fills the $\mathbb{R}^d$.
So, by the breadth-first search algorithm a family of hyperplanes, each
defining a halfspace, is generated, and the intersection of these halfspaces
forms the halfspace depth trimmed region
\citep[see also][]{PaindaveineS12a,PaindaveineS12b}.
\cite{PaindaveineS12a} state that the average complexity of their
algorithm is not worse than $O(n^d)$ and the worst-case complexity is of order
$O(n^{d+1})$.

Based on an idea similar to~\citet{HallinPS10}, \citet{LiuZ14a} compute the
halfspace depth exactly by using a breadth-first search algorithm to cover
$\mathbb{R}^d$ and QHULL to define the direction cones.
In a most recent article \citet{Liu14} suggests two more algorithms for the
exact computation of the halfspace depth, which seem to be very fast.
One of these algorithms, the so-called refined combinatorial
algorithm, can be seen as a special case of the framework developed in the current paper.

\subsection{Approximation of the halfspace depth}\label{ssec:approx}
Even with the fastest available algorithms the exact calculation of 
the halfspace depth is very elaborate and amounts to exponential complexity in 
$n$ and $d$. Therefore, one tries to
save computation expenses by approximating the depth.
Following~\citet{Dyckerhoff04}, the halfspace depth satisfies the weak projection
property, i.e., it is the smallest achievable depth on all one-dimensional
projections, and thus can be bounded from above by univariate depths.

\citet{RousseeuwS98}, when suggesting the algorithm computing the halfspace depth for
$d=3$, explored four algorithms differing in how the directions to project the data
are generated. They propose to take a random subset of: (1) all lines connecting
$\bmz$ and a point from $\bmX$, (2) all lines connecting two points from $\bmX$,
(3) all lines normal to hyperplanes based on $\bmz$ and $d-1$ pairwise
distinct points from $\bmX$, (4) all lines normal to hyperplanes based on $d$
pairwise distinct points from $X$, and claim the last variant to work best.
\citet{CuestaANR08} suggest to generate directions uniformly on $\mathbb{S}^{d-1}$. This
method proves to be useful in classification. \citet{AfshaniC09} present a
randomized data structure keeping the approximated depth value in some range of
deviations from its real value.

The latter work of \citet{ChenMW13} determines the number of tries needed to
achieve a required precision exploiting the third approximation method of
\citet{RousseeuwS98}. The authors also present its generalization by projecting
$\bmz$ and $\bmX$ onto affine spaces of dimension greater than one. They report
the approximated depth values to be exact in most of the experiments and the
approximation errors never to be larger than $2/n$.

For practical applications, especially in higher dimension,
approximate algorithms seem to be promising and may outperform exact algorithms
in terms of computation time. However, for assessing the quality of approximate 
algorithms, exact algorithms are needed. Since a thorough comparison of 
approximate algorithms would go beyond the scope of this article, we decided to 
concentrate only on exact algorithms. The study of approximate algorithms
will be the topic of a forthcoming article.

\subsection{Proposal}
In this paper, we suggest a theoretical framework for computing the halfspace
depth, which yields a whole class of algorithms. For $k\in\{1,\dots,d-1\}$,
consider a tuple of $k$ (out of $n$) data points, which together with $\bmz$
span an affine subspace of dimension $k$. For each such tuple, the data are 
projected onto the corresponding orthogonal complement, 
and the halfspace depth is
computed as the sum of the depth in these two orthogonal subspaces. Further,
for some fixed $k$, the halfspace depth is obtained as the minimum of 
the depths over all such tuples. All proposed algorithms are capable of dealing 
with data that are not in general position and even with ties.

In what follows, we first develop the theoretical framework in
Section~\ref{sec:theory}, leading to the main result stated in
Theorem~\ref{thm:4}, which yields the above mentioned class of algorithms.
Further, three algorithms for $k=1,d-2,d-1$ are presented in
Section~\ref{sec:algorithms}. In Section~\ref{sec:experiments}, we discuss
implementation issues and provide a speed comparison of the three algorithms
for data in general and non-general position. Some further ideas concerning
the application of the proposed techniques are discussed in
Section~\ref{sec:outlook}.

We will use the following notation. The number of elements of a set $I$ is
denoted by $\#I$ and the complement of a set $I$ by $I^c$. The linear
hull of some points $\bmx_1,\dots,\bmx_k\in\mathbb{R}^d$ is denoted by
$\sp(\bmx_1,\dots,\bmx_k)$. For a subspace $U$ of $\mathbb{R}^d$, we denote
the orthogonal complement of $U$ by $U^\perp$.

\section{Theory}\label{sec:theory}
The \emph{halfspace depth} of a point $\bmz\in\mathbb{R}^d$ w.r.t. $n$
data points $\bmx_1,\dots,\bmx_n\in\mathbb{R}^d$ is defined by
\[
\HD(\bmz\,|\,\bmx_1,\dots,\bmx_n)=\frac1n\min_{\bmp\ne\bmNull}\#\{i\forwhich \bmp'\bmx_i\ge\bmp'\bmz\}\,.
\]
The halfspace depth of a point $\bmz$ is therefore simply the minimum fraction
of data points $\bmx_1,\dots,\bmx_n$ contained in a closed halfspace containing
$\bmz$. For simplicity, we will also use the shorter notation
$\HD(\bmz\,|\,\bmX)$.

The halfspace depth is affine invariant, in particular it is location invariant.
Therefore,
\[
\HD(\bmz\,|\,\bmx_1,\dots,\bmx_n)=\HD(\bmNull\,|\,\bmx_1-\bmz,\dots,\bmx_n-\bmz)\,,
\]
which shows that we can restrict ourselves w.l.o.g. to the case that the halfspace
depth of the origin has to be computed. Further, it will be useful to consider
the integer version of the halfspace depth,
\[
\nHD(\bmz\,|\,\bmX)= n\cdot \HD(\bmz\,|\,\bmX)\in\mathbb{N}_0\,.
\]
In this section we will assume that $\bmNull\notin\{\bmx_1,\dots,\bmx_n\}$.
If some of the data points are equal to $\bmNull$, then these points are removed
from the data set and their number is simply added to the (integer) halfspace 
depth of the origin w.r.t. the remaining points.
We will further assume that $\sp(\bmx_1,\dots,\bmx_n)=\mathbb{R}^d$.
If this is not the case, i.e., if the data points are contained in some
$k$-dimensional subspace, $k<d$, then the data points are mapped to 
$\mathbb{R}^k$ by a linear transformation of rank $k$, and the algorithms are
then applied to the transformed data points.

Under the assumptions from above, the (integer) halfspace depth can be written as
\[
\nHD(\bmNull\,|\,\bmX)=\min_{\bmp\ne\bmNull}\#\{i\,|\,\bmp'\bmx_i\ge 0\}\,.
\]
A vector $\bmp\ne\bmNull$ is called \emph{optimal for the data set $\bmX$}
if
\[
\nHD(\bmNull\,|\,\bmX)=\#\{i\,|\,\bmp'\bmx_i\ge 0\}\,.
\]
We will use the following notation: If $\bmp\ne\bmNull$, then
\[
I_\bmp^+=\{i\forwhich \bmp'\bmx_i>0\}\,,\quad
I_\bmp^0=\{i\forwhich \bmp'\bmx_i=0\}\,,\quad
I_\bmp^-=\{i\forwhich \bmp'\bmx_i<0\}\,,
\]
and the corresponding cardinalities are denoted by
\[
n_\bmp^+=\#I_\bmp^+\,,\quad
n_\bmp^0=\#I_\bmp^0\,,\quad
n_\bmp^-=\#I_\bmp^-\,.
\]
With this notation
\[
\nHD(\bmNull\,|\,\bmX)
=\min_{\bmp\ne\bmNull}(n_\bmp^++n_\bmp^0)
=n - \max_{\bmp\ne\bmNull}n_\bmp^-\,.
\]
For a subset $I$ of indices, $\bmX_I$ denotes the data set $(\bmx_i)_{i\in I}$
of all data points with indices in $I$.
If the data are not in general position, then the linear hull of some
data points $\bmx_{i_1},\dots,\bmx_{i_k}$ may contain additional data points.
For a set $I=\{i_1,\dots,i_k\}$ of indices we denote by
\[
I^*=\{j\forwhich \bmx_j\in\sp(\bmx_{i_1},\dots,\bmx_{i_k})\}
\]
the set of all indices $j$ such that $\bmx_j$ is contained in the linear hull
of $\bmx_{i_1},\dots,\bmx_{i_k}$. Obviously, $I\subset I^*$. The cardinality
of $I^*$ is denoted by $n_{I^*}$.
\begin{proposition}
If $\bmp\ne0$ is optimal for $\bmX$,
then $I_{\bmp}^0=\emptyset$, i.e., no data points lie on the boundary of the
closed halfspace defined by $\bmp$.
\end{proposition}
\textbf{Proof:} Assume that $I_\bmp^0\ne\emptyset$. Then, there is an index
$i_0\in I_\bmp^0$, i.e., $\bmp'\bmx_{i_0}=0$. Now consider $\tilde\bmp
=\bmp-\lambda\bmx_{i_0}$ where $\lambda>0$. By choosing $\lambda$ sufficiently
small we can guarantee that $\tilde\bmp'\bmx_i<0$ for all indices $i\in I_\bmp^-$.
On the other hand $\tilde\bmp'\bmx_{i_0}=-\lambda\|\bmx_{i_0}\|^2<0$. Thus,
$i_0\in I_{\tilde\bmp}^-$.
Consequently, $n_{\tilde\bmp}^->n_\bmp^-$, which contradicts the optimality of $\bmp$.
So, $I_\bmp^0=\emptyset$.\qed

We will need the following lemma, whose proof can be found in the appendix.
\begin{lemma}\label{lem.1}
If $\bmp\ne\bmNull$ and $0\le \dim \sp(\bmX_{I_\bmp^0})=l<d-1$, then there exists
$\bmx_{i_0}\notin\sp(\bmX_{I_\bmp^0})$ and a vector $\tilde\bmp=\bmp+\bmq\ne
\bmNull$ with $\bmq\in\sp(\bmX_{I_\bmp^0},\bmx_{i_0})$ such that
\[
I_{\tilde\bmp}^0=\left[I_\bmp^0\union\{i_0\}\right]^*\,,\quad
I_\bmp^+\subset I_{\tilde\bmp}^+\union I_{\tilde\bmp}^0\,,\quad
I_\bmp^-\subset I_{\tilde\bmp}^-\union I_{\tilde\bmp}^0\,.
\]
Further, $\dim \sp(\bmX_{I_{\tilde\bmp}^0})=l+1$.
\end{lemma}
This means that if we change $\bmp$ to $\tilde\bmp$, then at least one of the data
points that were contained in one of the open halfspaces defined by $\bmp$
is now on the boundary hyperplane defined by $\tilde\bmp$, whereas no data
points did change sides.

\begin{proposition}\label{prop.1}
If $\bmp\ne0$ is optimal for $\bmX$,
then for every $k$,
$1\le k< d$, there are $k$ linearly independent data points $\bmx_{i_1},\dots,
\bmx_{i_k}$ and a vector $\tilde\bmp=\bmp+\bmq\ne\bmNull$ where
$\bmq\in\sp(\bmx_{i_1},\dots,\bmx_{i_k})$ such that
\[
I_{\tilde\bmp}^0=\{i_1,\dots,i_k\}^*\,,\quad
I_\bmp^+\subset I_{\tilde\bmp}^+\union I_{\tilde\bmp}^0\,,\quad
I_\bmp^-\subset I_{\tilde\bmp}^-\union I_{\tilde\bmp}^0\,.
\]
\end{proposition}
\textbf{Proof:} The proposition follows from repeated application of
Lemma~\ref{lem.1}.\qed

\begin{proposition}
Let $\bmx_{i_1},\dots,\bmx_{i_k}$ be linearly independent and $\bmp\ne\bmNull$
such that $I=\{i_1,\dots,i_k\}\subset I_\bmp^0$.
Then,
\[
\nHD(\bmNull\forwhich \bmX)\le
(n_\bmp^++n_\bmp^0)-\bigl(n_{I^*}-\nHD(\bmNull\forwhich\bmX_{I^*})\bigr)\,.
\]
\end{proposition}
\textbf{Proof:} Let $\bmNull\ne\bmq\in\sp(\bmx_{i_1},\dots,\bmx_{i_k})$ be
optimal for the data set $\bmX_{I^*}$.
Consider the direction $\tilde\bmp=\bmp+\lambda\bmq$ with $\lambda>0$. We have
\[
\tilde\bmp'\bmx_i=\bmp'\bmx_i+\lambda\bmq'\bmx_i\,.
\]
By choosing $\lambda$ sufficiently small we can guarantee that
$\tilde\bmp'\bmx_i$ and $\bmp'\bmx_i$ have the same sign for all $i\in I_\bmp^+
\union I_\bmp^-$.
For $\bmx_i\in\bmX_{I^*}$ it holds that $\tilde\bmp'\bmx_i=\lambda\bmq'\bmx_i$. Since
$\bmq$ is optimal for $\bmX_{I^*}$, there are $n_{I^*}-\nHD(\bmNull\forwhich\bmX_{I^*})$
data points in $\bmX_{I^*}$ for which $\bmq'\bmx_i<0$. Therefore,
\begin{align*}
\nHD(\bmNull\forwhich\bmX)
&\le\#\{i\,|\,\tilde\bmp'\bmx_i\ge 0\}\\
&=\#\{i\,|\,\bmp'\bmx_i\ge0\}-\#\{i\in I^*\,|\,\bmq'\bmx_i< 0\}\\
&=(n_\bmp^++n_\bmp^0)-\bigl(n_{I^*}-\nHD(\bmNull\forwhich\bmX_{I^*})\bigr)\,.
\end{align*}
\qed

\begin{proposition}
For every $k$, $1\le k< d$, there is a set $I=\{i_1,\dots,i_k\}$ of indices
and a vector $\tilde\bmp\ne\bmNull$ such that $\bmx_{i_1},\dots,\bmx_{i_k}$ are
linearly independent, $I_{\tilde\bmp}^0=I^*$, and
\[
\nHD(\bmNull\forwhich \bmX)
\ge(n_{\tilde\bmp}^++n_{\tilde\bmp}^0)
-\bigl(n_{I^*}-\nHD(\bmNull\forwhich \bmX_{I^*})\bigr)\,.
\]
\end{proposition}
\textbf{Proof:} Let $\bmp_0\ne\bmNull$ be optimal for $\bmX$.
Choose $\bmq$ and $\tilde\bmp=\bmp_0+\bmq$ as in
Proposition~\ref{prop.1}. For
$i\in I_{\bmp_0}^-\intersection I_{\tilde\bmp}^0$ it holds that
$\bmp_0'\bmx_i<0$ and $\tilde\bmp'\bmx_i=0$ which implies $\bmq'\bmx_i>0$ and thus
$(-\bmq)'\bmx_i<0$. Therefore,
\[
\nHD(\bmNull\forwhich \bmX_{I_{\tilde\bmp}^0})\le n_{\tilde\bmp}^0 - \#(I_{\bmp_0}^-\intersection I_{\tilde\bmp}^0)\,.
\]
Then,
\begin{align*}
\nHD(\bmNull\forwhich\bmX)
&=\#\{i\,|\,\bmp_0'\bmx_i>0\}\\
&=\#\{i\,|\,\tilde\bmp'\bmx_i\ge 0\}-\#(I_{\bmp_0}^-\intersection I_{\tilde\bmp}^0)\\
&\ge(n_{\tilde\bmp}^++n_{\tilde\bmp}^0)
-\bigl(n_{\tilde\bmp}^0-\nHD(\bmNull\forwhich \bmX_{I_{\tilde\bmp}^0})\bigr)\,.
\end{align*}
Because of Proposition~\ref{prop.1}, it holds that $I_{\tilde\bmp}^0=I^*$ 
and thus $n_{\tilde\bmp}^0=n_{I^*}$, which completes the proof.
\qed\\

For the following denote by $\mathcal{L}_k$ the set of all subsets $I$ of order
$k$ of $\{1,\dots,n\}$ such that the points $(\bmx_i)_{i\in I}$ are
linearly independent.
\begin{theorem}
For each $k$ such that $1\le k<d$ it holds that
\[
\nHD(\bmNull\forwhich \bmX)=\min_{I\in\mathcal{L}_k}
\left[
\left(\min_{\bmp\in \bmX_I^\perp\setminus\{\bmNull\}}\left(n_{\bmp}^++n_{\bmp}^0\right)\right)
-\Bigl(n_{I^*}-\nHD(\bmNull\forwhich\bmX_{I^*})\Bigr)
\right]\,.
\]
\end{theorem}
\textbf{Proof:} The proof follows immediately from the preceding two
propositions.\ \hfill\qed

We will now show how
\[
\min_{\bmp\in \bmX_I^\perp\setminus\{\bmNull\}}(n_{\bmp}^++n_{\bmp}^0)
\]
can be computed as the halfspace depth of a projection of the data points.

Let $I$ be a subset of $\{1,\dots,n\}$ of order $k$ such that the data points
$\bmx_i$, $i\in I$, are linearly independent. Let further
$\bma_1,\dots,\bma_{d-k}$ be a basis  of the orthogonal complement of
$\sp(\bmx_{i_1},\dots,\bmx_{i_k})$ and $\bmA_I$ the matrix whose columns are the
$\bma_i$. Thus, every vector $\bmp\in\bmX_I^\perp$ is a linear combination of
$\bma_1,\dots,\bma_{d-k}$, i.e., $\bmp=\bmA_I\tilde\bmp$ for some $\tilde\bmp\in
\mathbb{R}^{d-k}$,
and the map $\mathbb{R}^{d-k}\to \bmX_I^\perp$,
$\tilde\bmp\mapsto\bmA_I\tilde\bmp=:\bmp$,
is a bijection.
Since
\[
\tilde\bmp'(\bmA_I'\bmx_i)\ge0
\quad\iff\quad
(\tilde\bmp'\bmA_I')\bmx_i\ge0
\quad\iff\quad
\bmp'\bmx_i\ge0\,,
\]
we conclude
\begin{align*}
\min_{\bmp\in \bmX_I^\perp\setminus\{\bmNull\}}(n_{\bmp}^++n_{\bmp}^0)
&=\min_{\bmp\in \bmX_I^\perp\setminus\{\bmNull\}}\#\{i\forwhich\bmp'\bmx_i\ge0\}\\
&=\min_{\tilde\bmp\in\mathbb{R}^{d-k}\setminus\{\bmNull\}}
\#\{i\forwhich\tilde\bmp'(\bmA_I'\bmx_i)\ge0\}\\
&=\nHD(\bmNull\forwhich\bmA_I'\bmX)\,.
\end{align*}
A further simplification arises from the fact that all points $\bmx_i$,
$i\in I^*$, are mapped to the origin by $\bmA_I'$. Therefore, these points can
be removed from the data set and the halfspace depth is computed w.r.t. the
data set $\bmx_i$, $i\in (I^*)^c$.
Therefore,
\[
\nHD(\bmNull\forwhich\bmA_I'\bmX)=\nHD(\bmNull\forwhich\bmA_I'\bmX_{(I^*)^c})
+n_{I^*}\,,
\]
and finally
\begin{align*}
\nHD(\bmNull\forwhich \bmX)
&=\min_{I\in\mathcal{L}_k}
\left[
\nHD(\bmNull\forwhich\bmA_I'\bmX)-\bigl(n_{I^*}-\nHD(\bmNull\forwhich\bmX_{I^*})\bigr)
\right]\\
&=\min_{I\in\mathcal{L}_k}
\left[
\nHD(\bmNull\forwhich\bmA_I'\bmX_{(I^*)^c})+\nHD(\bmNull\forwhich\bmX_{I^*})
\right]\,. 
\end{align*}


In the same way as above it can be shown that
\[
\nHD(\bmNull\forwhich\bmX_{I^*})=\nHD(\bmNull\forwhich \bmP_I'\bmX_{I^*})\,,
\]
where $\bmP_I=[\bmx_{i_1},\dots,\bmx_{i_k}]$. Therefore, the following theorem
holds.

\begin{theorem}\label{thm:4}
With the notation from above, for each $1\le k< d$ it holds that
\[
\nHD(\bmNull\forwhich \bmX)=\min_{I\in\mathcal{L}_k}
\Bigl[
\nHD(\bmNull\forwhich\bmA_I'\bmX_{(I^*)^c})+\nHD(\bmNull\forwhich\bmP_I'\bmX_{I^*})
\Bigr]\,.
\]
\end{theorem}
Note that for each subset $I$ of $k$ linearly independent data points the data
points fall in one of two categories: The points whose projections on the
orthogonal complement of $\sp(\bmX_{I})$ are different from $\bmNull$ and those
who are equal to $\bmNull$. The former points are taken into account by
$\nHD(\bmNull\forwhich\bmA_I'\bmX_{(I^*)^c})$, whereas the latter ones are
considered by $\nHD(\bmNull\forwhich\bmP_I'\bmX_{I^*})$. Usually, there will be
much more points of the first category than of the second one. The computation
of $\nHD(\bmNull\forwhich\bmA_I'\bmX_{(I^*)^c})$ is done in dimension $d-k$,
whereas the computation of $\nHD(\bmNull\forwhich\bmP_I'\bmX_{I^*})$ is
done in dimension $k$. Thus, by the preceding theorem the calculation of one
depth value in $d$-space is reduced to calculating many depth values in
$(d-k)$-space (and in $k$-space). By choosing $k$ we can control how much the
dimension is reduced in each step. The price for a higher dimension reduction
is that more subsets $I$ have to be considered in that step.

An important special case of the algorithm arises when the data points
$\bmx_1,\dots,\bmx_n$ and $\bmNull$ are in general position. In that case,
any linear subspace of dimension $k$, $1\le k<d$, contains at most $k$ of
the data points $\bmx_1,\dots,\bmx_n$. Therefore, if $\#I=k$, then $I^*=I$ and
$n_{I^*}=k$. Further, $\nHD(\bmNull\forwhich\bmP_I'\bmX_{I^*})=0$, which can be seen by
choosing $\bmp$ such that $\bmp'(\bmP_I'\bmx_{i_1}-\bmP_I'\bmx_{i_j})=0$,
$j=2,\dots,k$. Therefore, we get the following corollary of the above theorem.

\begin{corollary}
\label{cor:gp}
If the data points $\bmx_1,\dots,\bmx_n$ and $\bmNull$ are in general 
position, then for each $1\le k< d$ it holds that
\[
\nHD(\bmNull\forwhich \bmX)=\min_{I\in\mathcal{L}_k}
\nHD(\bmNull\forwhich\bmA_I'\bmX_{I^c})\,.
\]
\end{corollary}

\section{Algorithms}\label{sec:algorithms}
The result of the previous section gives rise to several algorithms.
In these algorithms the dimensionality of the data is reduced at different
rates. When the dimension is reduced to $d=1$ or $d=2$, specialized algorithms
may be used. For the case $d=1$ the standard algorithm of complexity $O(n)$ is
used. For bivariate data the algorithm of \cite{RousseeuwR96} with complexity
$O(n\log n)$ (or any other algorithm with this complexity) may be used.

\subsection{Combinatorial algorithm, $k=d-1$}
\begin{figure}[ht]
\begin{center}
    \includegraphics[keepaspectratio=true,scale=1]{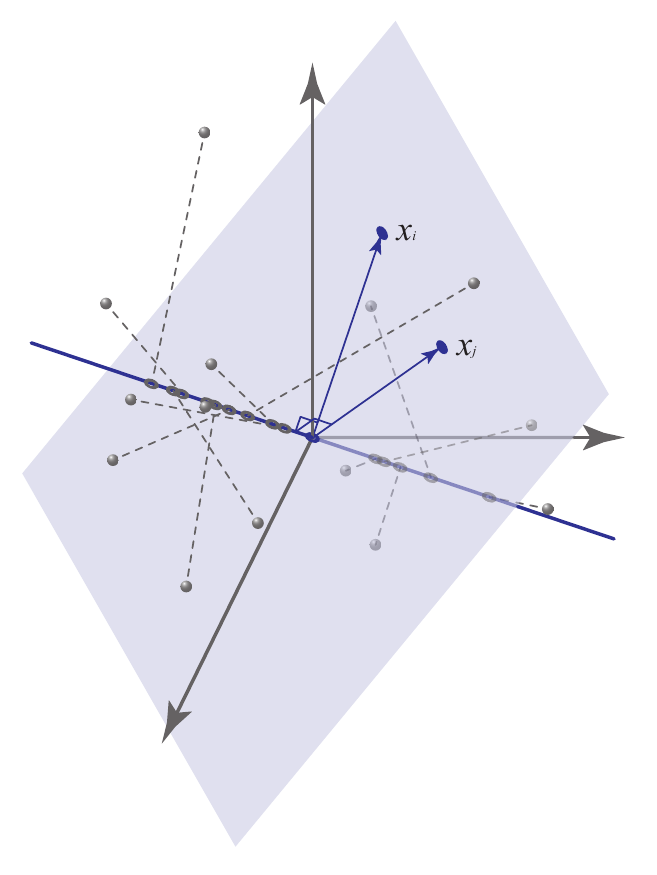}\,
    \includegraphics[keepaspectratio=true,scale=1]{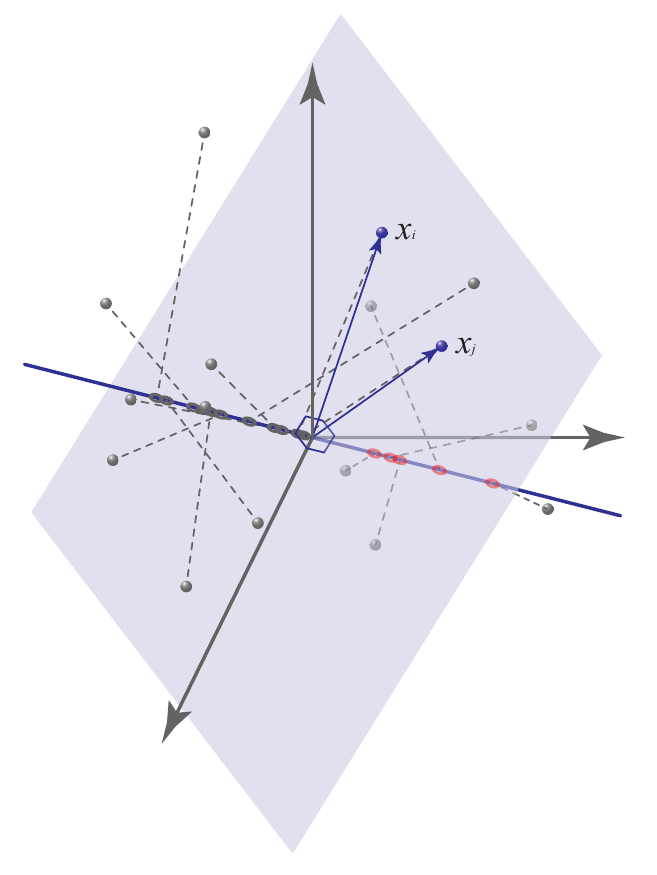}
	\caption{Illustration of the combinatorial algorithm, $k=d-1$}
    \label{fig:piccmb}
\end{center}
\end{figure}

If we choose $k=d-1$, this results in the so-called \emph{combinatorial
algorithm} (Algorithm~\ref{alg:d-1}).
In this algorithm, all hyperplanes defined by $d-1$ linearly independent data
points are considered. For each such hyperplane the data are projected in the
direction normal to the hyperplane. Thus, the dimensionality is in one step
reduced to dimension one. Only if there are more than $d-1$ data points in the
considered hyperplane (which can only occur when the data are not in general
position), then for these data points $\bmy_i=\bmP_I'\bmx_i$ is calculated and
the procedure $\nHD$ is recursively called for the data points
$\bmy_{j_1},\dots,\bmy_{j_l}$.
The algorithm is illustrated in Figure~\ref{fig:piccmb}.

In the case of data in general position, the algorithm will never enter the
recursion. Then, for each processed hyperplane the complexity of this algorithm
is of order $O(n)$. Since there are $\binom{n}{d-1}$ subsets of $d-1$ data
points, the overall complexity of the algorithm is $\binom{n}{d-1}O(n)=O(n^d)$.

\begin{algorithm}[ht]
\caption{Combinatorial algorithm, $k=d-1$}\label{alg:d-1}
\begin{algorithmic}[1]
\Function{nHD\_Comb}{$d,\bmx_1,\dots,\bmx_n$}  \Comment{Halfspace depth of $0$}
\If{$d=1$} \Return \Call{nHD1}{$\bmx_1,\dots,\bmx_n$}\EndIf
\State $n_{min}\gets n$
\For{each subset $I\subset\{1,\dots,n\}$ of order $d-1$}
\If{$(\bmx_i)_{i\in I}$ linearly independent}
\State Compute $\bmp_I$ such that $\bmp_I'\bmx_i=0$ for all $i\in I$
\ForAll{$\bmx_j$}
\State $z_j\gets \bmp_I'\bmx_j$ \Comment project data points in direction $\bmp_I$
\EndFor
\State $n_{new}\gets \min\Bigl\{\#\{z_j>0\},\#\{z_j<0\}\Bigr\}$
\If{$\#\{z_j=0\}>d-1$}
\State $\bmP_I\gets\mathrm{Matrix}[(\bmx_i)_{i\in I}]$
\ForAll{indices $j$ with $z_j=0$}
\State $\bmy_j\gets \bmP_I'\bmx_{j}$
\EndFor
\State $n_{new}\gets n_{new}+\text{\Call{nHD\_Comb}{$d-1,\bmy_{j_1},\dots,\bmy_{j_l}$}}$
\EndIf
\If{$n_{new}<n_{min}$} $n_{min}\gets n_{new}$\EndIf
\EndIf
\EndFor
\State\Return $n_{min}$
\EndFunction
\end{algorithmic}
\end{algorithm}

\subsection{Combinatorial algorithm, $k=d-2$}
Another possibility is to use $k=d-2$ (Algorithm~\ref{alg:d-2}).
In that case, the data points are
directly projected into the 2-dimensional space. This has the advantage that
for the projected points the algorithm of \cite{RousseeuwR96} for the
bivariate halfspace depth can be used. Since this algorithm has a complexity
of $O(n\log n)$ and there are $\binom{n}{d-2}$ subsets of order $d-2$, the
complexity of this algorithm is of order
$\binom{n}{d-2}O(n\log n)=O(n^{d-1}\log n)$. Thus, this algorithm has a better
complexity than the naive combinatorial algorithm with $k=d-1$.
This combinatorial algorithm has been independently proposed in a manuscript
by \cite{Liu14}.

\begin{algorithm}[ht]
\caption{Combinatorial algorithm, $k=d-2$}\label{alg:d-2}
\begin{algorithmic}[1]
\Function{nHD\_Comb2}{$d,\bmx_1,\dots,\bmx_n$}  \Comment{Halfspace depth of $0$}
\If{$d=1$} \Return \Call{nHD1}{$\bmx_1,\dots,\bmx_n$}\EndIf
\If{$d=2$} \Return \Call{nHD2}{$\bmx_1,\dots,\bmx_n$}\EndIf
\State $n_{min}\gets n$
\For{each subset $I\subset\{1,\dots,n\}$ of order $d-2$}
\If{$(\bmx_i)_{i\in I}$ linearly independent}
\State Compute a basis $\bma_{1},\bma_{2}$ of the orthogonal complement of $(\bmx_i)_{i\in I}$
\State $\bmA_I\gets\mathrm{Matrix}[\bma_{1},\bma_{2}]$
\State $\bmP_I\gets\mathrm{Matrix}[\bmx_{i_1},\dots,\bmx_{i_{d-2}}]$
\ForAll{$\bmx_j$}
\If{$\bmA_I'\bmx_j\ne\bmNull$} $\bmy_j\gets\bmA_I'\bmx_j$\EndIf
\State \hphantom{\textbf{if} $\bmA_I'\bmx_i\ne\bmNull$}\ \ \textbf{else}\ \ $\bmz_j\gets \bmP_I'\bmx_{j}$
\EndFor
\State $l\gets \#\{j:\bmA_I'\bmx_j\ne\bmNull\}$
\State $n_{new}\gets \text{\Call{nHD2}{$\bmy_{j_1},\dots,\bmy_{j_l}$}}$
\If{$n-l>d-2$}
\State $n_{new}\gets n_{new}+\text{\Call{nHD\_Comb2}{$d-2,\bmz_{j_1},\dots,\bmz_{j_{n-l}}$}}$
\EndIf
\If{$n_{new}<n_{min}$} $n_{min}\gets n_{new}$\EndIf
\EndIf
\EndFor
\State\Return $n_{min}$
\EndFunction
\end{algorithmic}
\end{algorithm}

\subsection{Recursive algorithm, $k=1$}
\begin{figure}[ht]
\begin{center}
    \includegraphics[keepaspectratio=true,scale=0.75]{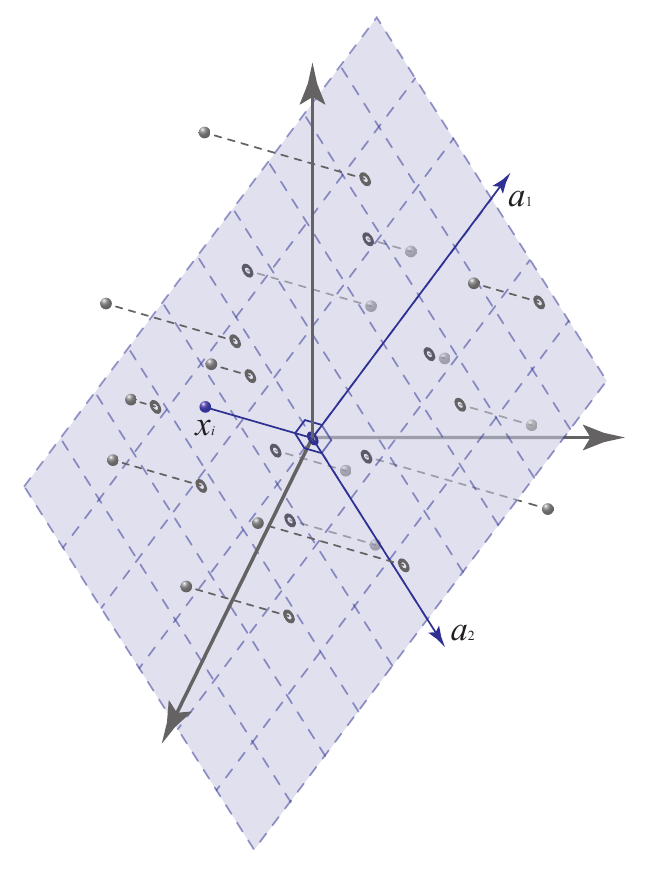}\,
    \includegraphics[keepaspectratio=true,scale=0.65]{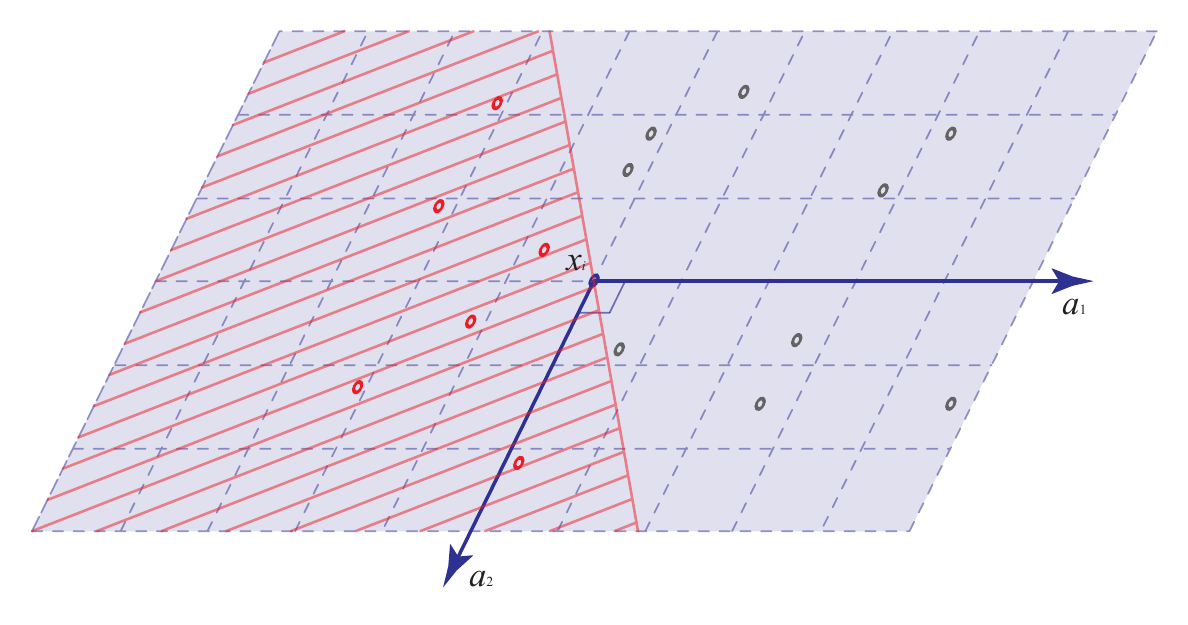}
	\caption{Illustration of the recursive algorithm, $k=1$}
    \label{fig:picrec}
\end{center}
\end{figure}

The other extreme is the case, where we choose $k=1$. This yields the so-called
\emph{recursive algorithm} (Algorithm~\ref{alg:1}).
In this algorithm, in the outer loop all data points
$\bmx_i$ are considered, and the data are projected on the hyperplane orthogonal
to $\bmx_i$. For the projected data points (with the exception of the data
points that are a multiple of $\bmx_i$ and are thus mapped to the origin), the
algorithm is called recursively. Thus, in each step the dimensionality is
reduced only by one. The recursion stops when $d=2$, in which case the algorithm
of \cite{RousseeuwR96} is applied. Note that the recursive algorithm can be
viewed as a generalization of the algorithm for the case $d=3$ in
\cite{RousseeuwS98}. Figure~\ref{fig:picrec} shows an illustration of the recursive
algorithm.

In the recursive algorithm the depth w.r.t. $d$-variate data is computed as the
minimum over $n$ depths w.r.t. $(d-1)$-variate data. Therefore, the complexity
for $d$-variate data is $n$ times the complexity for $(d-1)$-variate data.
Since the recursion is stopped when $d=2$, in which case the
$O(n\log n)$-algorithm of \cite{RousseeuwR96} is used, this results in an
overall complexity of $n^{d-2}\,O(n\log n)=O(n^{d-1}\log n)$. Note that this
remains true even if the data are not in general position.

\begin{algorithm}[ht]
\caption{Recursive algorithm, $k=1$}\label{alg:1}
\begin{algorithmic}[1]
\Function{nHD\_Rec}{$d,\bmx_1,\dots,\bmx_n$}  \Comment{Halfspace depth of $0$}
\If{$d=1$} \Return \Call{nHD1}{$\bmx_1,\dots,\bmx_n$}\EndIf
\If{$d=2$} \Return \Call{nHD2}{$\bmx_1,\dots,\bmx_n$}\EndIf
\State $n_{min}\gets n$
\ForAll{$\bmx_i$}
\State Compute a basis $\bma_{1},\dots,\bma_{d-1}$ of the hyperplane with normal $\bmx_i$
\State $\bmA_I\gets\mathrm{Matrix}[\bma_{1},\dots,\bma_{d-1}]$
\ForAll{$\bmx_j$}
\If{$\bmA_I'\bmx_j\ne\bmNull$} $\bmy_j\gets\bmA_I'\bmx_j$\EndIf
\State \hphantom{\textbf{if} $\bmA_I'\bmx_i\ne\bmNull$}\ \ \textbf{else}\ \ $z_j\gets \bmx_i'\bmx_{j}$
\EndFor
\State $l\gets \#\{j:\bmA_I'\bmx_j\ne\bmNull\}$
\State $n_{new}\gets \text{\Call{nHD\_Rec}{$d-1,\bmy_{j_1},\dots,\bmy_{j_l}$}}$
\State \hfill$+\min\Bigl\{\#\{z_j>0\},\#\{z_j<0\}\Bigr\}$
\If{$n_{new}<n_{min}$} $n_{min}\gets n_{new}$\EndIf
\EndFor
\State\Return $n_{min}$
\EndFunction
\end{algorithmic}
\end{algorithm}
\clearpage
\FloatBarrier

\section{Implementation and experiments}\label{sec:experiments}

All three variants of the algorithm, regarded in the previous section, have
been implemented in C++, without using external libraries, thus maintaining
maximum inter-system portability. The source code can be downloaded
from the website
\url{http://www.wisostat.uni-koeln.de/dyckerhoff.html?&L=1}
or directly from
\url{http://www.wisostat.uni-koeln.de/sites/statistik/source/HD_Dyckerhoff_Mozharovskyi.zip}.
Additionally, the implementation has been interfaced to the \texttt{R}-environment in the package \texttt{ddalpha}.
For any $k$, the algorithm can be easily implemented in any programming environment.
For $k=1,...,d-2$ the calculation of the orthogonal complement (which can
easily be done, e.g., using the Gauss-Jordan method) and the routine
\texttt{nHD2} by \citet{RousseeuwR96}
have to be implemented. 
For $k=d-1$ even the \texttt{nHD2}-routine is of no need.

The algorithms in the preceding section should not be called directly, but rather
be included in a wrapping function, which does the necessary preprocessing of the
data. In the preprocessing, the point $\bmz$ is first subtracted from the
data so that the halfspace depth of the origin has to be calculated.
Second, all data points which are equal to the origin are removed from the data.
Their number is stored and later added to the result of \texttt{nHD}.
Third, if the data points are contained in some $k$-dimensional 
subspace, $k<d$, then the data points are mapped to $\mathbb{R}^k$ as described 
in Section~\ref{sec:theory}.
In an optional fourth step, the data could be scaled to have a norm of 
one. This does not change the halfspace depth, but has the advantage that all 
data have the same order of magnitude, which should reduce numerical problems.

All algorithms based on Theorem~\ref{thm:4} can be further improved by exiting
the main loop as soon as $n_{min}$ drops to zero, since in that case no further
improvement is possible. However, this speed-up is data dependent and
occurs only if the origin is outside the convex hull of the data. Therefore, we
choose not to incorporate this modification into the algorithms that we used in
our experiments (see below) to get stable computation times.

Due to the independent repetition of similar operations, the algorithms based
on Theorem~\ref{thm:4} (for different values of $k$)
possess high parallelization abilities, which grow with $k$.                       
Clearly, for data in general position, the complexity of the algorithms for
$1\le k\le d-2$ is $O(n^{d-1}\log n)$ and is $O(n^d)$ for $k=d-1$.
However, the exact execution time depends on the implementation of the single
steps, routines, memory structures, etc.,
and can differ in practice from the values reported in Tables~\ref{tab1} and
\ref{tab2}.                                                                         
As we will see later in this section, each of the considered algorithms can
show the best results (compared to the remaining two algorithms) for proper
constellations of $n$ and $d$. The algorithms' performance may differ a lot
depending on whether the data are in general position or not.
In all experiments, we used one kernel of the Intel Core~i7-2600~(3.4~GHz)
processor having enough physical memory.

First, we consider data $\bmX$ drawn randomly from a multivariate standard
normal distribution $N\bigl(\mathbf{0}_d,\mathbf{I}_d\bigr)$, where the
depth of the origin w.r.t.\ $\bmX$ is computed.
Table~\ref{tab1} presents the execution times of the algorithms in seconds
(in each cell the upper, middle, and lower lines correspond to $k=1,2,d-1$
respectively), averaged over $10$ tries. Such a small number of tries is
sufficient, as the execution times are extremely stable for all chosen values of
$n$ and $d$ and differ by a few percents only. We vary $d$ from $3$ to $10$ and
increase $n=10\cdot 2^i$ with $i=2,3,...$ till
the execution time exceeds one hour.
As one can see, because of the recurrent structure, for fixed $n$ and with
increasing $d$, the algorithm with $k=1$ is outperformed by $k=d-2$, which is
further outperformed by $k=d-1$. On the other hand, for fixed $d$ and with
increasing $n$, the algorithm with $k=d-1$ is outperformed by $k=1$ and $k=d-2$
because of the better complexity of the latter algorithms. Comparing the
algorithms with $k=1$ and with $k=d-2$ the former one is superior for dimension
$d=3$, whereas the latter performs better when $d>3$.

The designed framework allows for handling data for which the general position
assumption is violated. The results for $\bmX$ distributed uniformly on
$\{-2, -1, 0, 1, 2\}^d$ are presented in Table~\ref{tab2}. As mentioned above
(see also Corollary~\ref{cor:gp}), for data in general position, if $k=d-1$ or
$k=d-2$, no recursion is involved.
If $k=d-1$, for data in non-general position, the recursive calls can increase
the execution time and, in general, the algorithmic complexity.
Additionally, if $n$ is not large enough (depending on $d$), the computation
times depend heavily on the exact position of the points in $\mathbb{R}^d$ and
become unstable. Therefore, for the algorithm with $k=d-1$ the median was taken
instead of the mean when reporting the execution times for $d=7,...,10$.
The same effect occurs in the case $k=d-2$ as well, but the application of the
two-dimensional routine (nHD2) designed by~\cite{RousseeuwR96} seems to
compensate this increase in time by a quick handling of ties, especially when
$n$ gets larger.                                                                    
On the other hand, if $k=1$, ties are rather an advantage, and the
execution time of the algorithm decreases.

\begin{sidewaystable}[!h]
\caption{Execution times for data in general position, distributed as
$N\bigl(\mathbf{0}_d,\mathbf{I}_d\bigr)$, averaged over 10 tries, in seconds
(three significant digits).
Variants of the algorithm with $k=1,d-2,d-1$ are presented in the first, second,
and third rows of each cell, respectively.}
\label{tab1}
{\scriptsize
\centerline{\begin{tabular}{rrrrrrrrrrrrrrr}
\toprule
& $n=40$ & 80 & 160 & 320 & 640 & 1280 & 2560 & 5120 & 10240 & 20480 & 40960 & 81920 \\ \midrule
$d=3$ & 0.000 & 0.000 & 0.000 & 0.011 & 0.047 & 0.184 & 0.780 & 3.19 & 13.2 & 54.5 & 225 & 933 \\
 & 0.002 & 0.002 & 0.003 & 0.016 & 0.063 & 0.250 & 1.03 & 4.22 & 17.4 & 72.2 & 293 & 1210  \\
 & 0.000 & 0.003 & 0.014 & 0.117 & 0.936 & 7.60 & 61.3 & 519 & --- & --- & --- & --- \\ \midrule
4 & 0.006 & 0.048 & 0.402 & 3.36 & 28.2 & 235 & 1960 & --- & --- & --- & --- & --- \\
 & 0.005 & 0.038 & 0.302 & 2.50 & 20.4 & 166 & 1360 & --- & --- & --- & --- & --- \\
 & 0.005 & 0.055 & 0.784 & 12.3 & 203 & 3290 & --- & --- & --- & --- & --- & --- \\ \midrule
5 & 0.205 & 3.62 & 62.5 & 1070 & --- & --- & --- & --- & --- & --- & --- & --- \\
 & 0.055 & 0.952 & 16.1 & 269 & --- & --- & --- & --- & --- & --- & --- & --- \\
 & 0.047 & 1.24 & 35.7 & 1110 & --- & --- & --- & --- & --- & --- & --- & --- \\ \midrule
6 & 7.32 & 275 & --- & --- & --- & --- & --- & --- & --- & --- & --- & --- \\
 & 0.506 & 18.4 & 633 & --- & --- & --- & --- & --- & --- & --- & --- & --- \\
 & 0.392 & 21.6 & 1250 & --- & --- & --- & --- & --- & --- & --- & --- & --- \\ \midrule
7 & 257 & --- & --- & --- & --- & --- & --- & --- & --- & --- & --- & --- \\
 & 3.60 & 278 & --- & --- & --- & --- & --- & --- & --- & --- & --- & --- \\
 & 2.66 & 305 & --- & --- & --- & --- & --- & --- & --- & --- & --- & --- \\ \midrule
8 & --- & --- & --- & --- & --- & --- & --- & --- & --- & --- & --- & --- \\
 & 21.4 & 3550 & --- & --- & --- & --- & --- & --- & --- & --- & --- & --- \\
 & 14.3 & 3470 & --- & --- & --- & --- & --- & --- & --- & --- & --- & --- \\ \midrule
9 & --- & --- & --- & --- & --- & --- & --- & --- & --- & --- & --- & --- \\
 & 107 & --- & --- & --- & --- & --- & --- & --- & --- & --- & --- & --- \\
 & 69.8 & --- & --- & --- & --- & --- & --- & --- & --- & --- & --- & --- \\ \midrule
10 & --- & --- & --- & --- & --- & --- & --- & --- & --- & --- & --- & --- \\
 & 439 & --- & --- & --- & --- & --- & --- & --- & --- & --- & --- & --- \\
 & 289 & --- & --- & --- & --- & --- & --- & --- & --- & --- & --- & --- \\ \bottomrule
\end{tabular}}
}
\end{sidewaystable}

\begin{sidewaystable}[!h]
\caption{Execution times for data in non-general position, distributed as
$U(\{-2, -1, 0, 1, 2\}^d)$, averaged over 10 tries, in seconds
(three significant digits). Variants
of the algorithm with $k=1,d-2,d-1$ are presented in the first, second, and
third rows of each cell, respectively. For $k=d-1$ and $d=7,...,10$ the median
is reported.}
\label{tab2}

{\scriptsize
\centerline{\begin{tabular}{rrrrrrrrrrrrrr}
\toprule
& $n=40$ & 80 & 160 & 320 & 640 & 1280 & 2560 & 5120 & 10240 & 20480 & 40960 & 81920 & 163840 \\ \midrule
$d=3$ & 0.000 & 0.002 & 0.002 & 0.009 & 0.036 & 0.139 & 0.530 & 2.11 & 8.33 & 33.1 & 132 & 530 & 2290 \\
 & 0.000 & 0.002 & 0.005 & 0.013 & 0.052 & 0.198 & 0.773 & 3.14 & 12.6 & 49.9 & 195 & 786 & 3480 \\
 & 0.002 & 0.019 & 0.188 & 2.11 & 26.1 & 384 & --- & --- & --- & --- & --- & --- & --- \\ \midrule
4 & 0.005 & 0.045 & 0.375 & 3.06 & 24.6 & 196 & 1560 & --- & --- & --- & --- & --- & --- \\
 & 0.005 & 0.036 & 0.291 & 2.35 & 18.5 & 141 & 1110 & --- & --- & --- & --- & --- & --- \\
 & 0.203 & 5.47 & 235 & --- & --- & --- & --- & --- & --- & --- & --- & --- & --- \\ \midrule
5 & 0.202 & 3.56 & 60.5 & 1020 & --- & --- & --- & --- & --- & --- & --- & --- & --- \\
 & 0.063 & 1.06 & 17.4 & 283 & --- & --- & --- & --- & --- & --- & --- & --- & --- \\
 & 4.49 & 869 & --- & --- & --- & --- & --- & --- & --- & --- & --- & --- & --- \\ \midrule
6 & 7.29 & 272 & --- & --- & --- & --- & --- & --- & --- & --- & --- & --- & --- \\
 & 0.570 & --- & --- & --- & --- & --- & --- & --- & --- & --- & --- & --- & --- \\
 & 78.1 & --- & --- & --- & --- & --- & --- & --- & --- & --- & --- & --- & --- \\ \midrule
7 & 256 & --- & --- & --- & --- & --- & --- & --- & --- & --- & --- & --- & --- \\
 & 4.34 & 315 & --- & --- & --- & --- & --- & --- & --- & --- & --- & --- & --- \\
 & 227 & --- & --- & --- & --- & --- & --- & --- & --- & --- & --- & --- & --- \\ \midrule
8 & --- & --- & --- & --- & --- & --- & --- & --- & --- & --- & --- & --- & --- \\
 & 24.2 & --- & --- & --- & --- & --- & --- & --- & --- & --- & --- & --- & --- \\
 & 754 & --- & --- & --- & --- & --- & --- & --- & --- & --- & --- & --- & --- \\ \midrule
9 & --- & --- & --- & --- & --- & --- & --- & --- & --- & --- & --- & --- & --- \\
 & 144 & --- & --- & --- & --- & --- & --- & --- & --- & --- & --- & --- & --- \\
 & 1750 & --- & --- & --- & --- & --- & --- & --- & --- & --- & --- & --- & --- \\ \midrule
10 & --- & --- & --- & --- & --- & --- & --- & --- & --- & --- & --- & --- & --- \\
 & 457 & --- & --- & --- & --- & --- & --- & --- & --- & --- & --- & --- & --- \\
 & 1800 & --- & --- & --- & --- & --- & --- & --- & --- & --- & --- & --- & --- \\ \bottomrule
\end{tabular}}
}
\end{sidewaystable}

\section{Conclusions}\label{sec:outlook}
In this paper a class of combinatorial algorithms is presented which calculate
the halfspace depth of $\bm0$ w.r.t. $\bmX$ as the minimum over all combinations
of $k$ points, $k\in\{1,2,...,d-1\}$.
For each combination, the depth is calculated as the sum of the depths of points
lying in the affine space, spanned by these $k$ points and the origin, and its
orthogonal complement.
For each $k$, the algorithm can be easily implemented in any programming
environment.
None of the algorithms requires the data to 
be in general position or that
the data have to be perturbed.
For a given hardware and under the assumption of general position (or negligible
violation of it), the computation times are stable for chosen $d$ and $n$, and
thus are predictable.

The empirical study shows that for each of the presented algorithms a pair
$(d,n)$ can be found where one algorithm outperforms the others in terms of
speed.
This suggests the development of a hybrid algorithm, which will choose $k$
depending on $d$ and $n$.
This hybrid algorithm should be tuned for the corresponding implementation and
hardware.

The developed framework may be extended to calculate further depths of
combinatorial nature.
Subsampling on the entire set of combinations for some (not necessarily equal)
$k$'s may be used for computation of approximate depth values.

\clearpage

\appendix

\section{Proof of Lemma~\ref{lem.1}}
Denote by $\bmy_i$ the projection of
$\bmx_i$ onto $\sp(\bmX_{I_\bmp^0})^\perp$, the orthogonal complement of
$\sp(\bmX_{I_\bmp^0})$. Then, every data point can be uniquely represented as
$\bmx_i=\bmy_i+\bmd_i$, where $\bmy_i\in\sp(\bmX_{I_\bmp^0})^\perp$ and
$\bmd_i\in\sp(\bmX_{I_\bmp^0})$. Note that $\bmp'\bmx_i=\bmp'\bmy_i$.
Now, let
\[
i_0=\argmin\left\{\frac{|\bmp'\bmy_i|}{\|\bmp\|\|\bmy_i\|}\forwhich i\in I_\bmp^+\union I_\bmp^-\right\}
\]
and
\[
\epsilon=\frac{\bmp'\bmy_{i_0}}{\|\bmp\|\|\bmy_{i_0}\|}=\cos\alpha(\bmp,\bmy_{i_0})\,,
\]
where $\alpha(\bmp,\bmy_{i_0})$ denotes the angle between $\bmp$ and $\bmy_{i_0}$.
Now, define $\tilde\bmp$ by
\[
\tilde\bmp=\bmp-\epsilon\frac{\|\bmp\|}{\|\bmy_{i_0}\|}\bmy_{i_0}\,.
\]
Since $\bmy_{i_0}=\bmx_{i_0}-\bmd_{i_0}\in\sp(\bmx_{i_0},\bmX_{I_\bmp^0})$,
it holds that $\tilde\bmp=\bmp+\bmq$, where $\bmq\in\sp(\bmx_{i_0},\bmX_{I_\bmp^0})$.
Further,
\[
\tilde\bmp'\bmx_i
=\tilde\bmp'(\bmy_i+\bmd_i)
=\bmp'\bmy_i+\bmp'\bmd_i-\epsilon\frac{\|\bmp\|}{\|\bmy_{i_0}\|}\bmy_{i_0}'\bmy_i
-\epsilon\frac{\|\bmp\|}{\|\bmy_{i_0}\|}\bmy_{i_0}'\bmd_i\,.
\]
Since $\bmd_i\perp\bmp$ and $\bmd_i\perp\bmy_i$, this reduces to
\[
\tilde\bmp'\bmx_i
=\bmp'\bmy_i-\epsilon\frac{\|\bmp\|}{\|\bmy_{i_0}\|}\bmy_{i_0}'\bmy_i\,.
\]
Now, consider the following cases:

If $i\in I_\bmp^+$, then
\[
\tilde\bmp'\bmx_i
=\bmp'\bmy_i-\epsilon\frac{\|\bmp\|}{\|\bmy_{i_0}\|}\bmy_{i_0}'\bmy_i
\ge 0
\]
since $\bmp'\bmy_i\ge|\epsilon|\|\bmp\|\|\bmy_i\|$ and
$|\bmy_{i_0}'\bmy_i|\le\|\bmy_{i_0}\|\|\bmy_i\|$.
Thus, $I_\bmp^+\subset I_{\tilde\bmp}^+\union I_{\tilde\bmp}^0$.

If $i\in I_\bmp^-$, then
\[
\tilde\bmp'\bmx_i
=\bmp'\bmy_i-\epsilon\frac{\|\bmp\|}{\|\bmy_{i_0}\|}\bmy_{i_0}'\bmy_i
\le 0
\]
since $\bmp'\bmy_i\le-|\epsilon|\|\bmp\|\|\bmy_i\|$ and
$|\bmy_{i_0}'\bmy_i|\le\|\bmy_{i_0}\|\|\bmy_i\|$.
Thus, $I_\bmp^-\subset I_{\tilde\bmp}^-\union I_{\tilde\bmp}^0$.

To conclude the proof, we now show that
$I_{\tilde\bmp}^0=\left[I_\bmp^0\union\{i_0\}\right]^*$.
First, note that
\begin{align*}
\tilde\bmp'\bmx_i
&=\bmp'\bmy_i-\epsilon\frac{\|\bmp\|}{\|\bmy_{i_0}\|}\bmy_{i_0}'\bmy_i\\
&=\|\bmp\|\|\bmy_i\|\cos\alpha(\bmp,\bmy_i)
-\epsilon\frac{\|\bmp\|}{\|\bmy_{i_0}\|}\|\bmy_{i_0}\|\|\bmy_i\|\cos\alpha(\bmy_{i_0},\bmy_i)\\
&=\|\bmp\|\|\bmy_i\|\left[\cos\alpha(\bmp,\bmy_i)
-\cos\alpha(\bmp,\bmy_{i_0})\cos\alpha(\bmy_{i_0},\bmy_i)\right]\,.
\end{align*}
The term in brackets is zero if and only if
\[
\frac{\cos\alpha(\bmp,\bmy_i)}{\cos\alpha(\bmp,\bmy_{i_0})}
=\cos\alpha(\bmy_{i_0},\bmy_i)\,.
\]
From the definition of $\epsilon$, the absolute value of the left hand side
is at least one. So, the only possible solution is if $\bmy_i$ is a
multiple of $\bmy_{i_0}$. Thus, $\tilde\bmp'\bmx_i=0$ if and only if $\bmy_i$
is a multiple of $\bmy_{i_0}$, i.e.,
\[
\bmx_i=\lambda\bmy_{i_0}+\bmd_i\in\sp(\bmy_{i_0},\bmX_{I_\bmp^0})\,.
\]
Therefore, $I_{\tilde\bmp}^0=\left[\{i_0\}\union I_\bmp^0\right]^*$,
as stated.
Since $\bmX_{I_{\tilde\bmp}^0}$ is generated from $\bmX_{I_{\bmp}^0}$ by
adjoining $\bmx_{i_0}$ (and maybe some other data points which are already in
the linear hull of $\bmX_{I_{\tilde\bmp}^0}\union\{\bmx_{i_0}\}$) its dimension
equals $l+1$, which completes the proof of the lemma.\qed

\section*{Acknowledgment}
We thank four anonymous referees for their careful reading and highly valuable
suggestions concerning the presentation of the paper.



\bigskip

\end{document}